\documentclass{mn2e}
\usepackage{epsfig}
\usepackage{epsf}
\usepackage{graphicx}

\title[Near-IR Spectroscopy of SN 2009ip]{Near-Infrared Spectroscopy
  of SN~2009ip's 2012 Brightening Reveals a Dusty Pre-Supernova
  Environment}

\author[Smith et al.]{Nathan Smith$^1$\thanks{Email:
    nathans@as.arizona.edu}, Jon C.\ Mauerhan$^1$, Mansi M.\
  Kasliwal$^2$, \& Adam J.\ Burgasser$^3$ \\ $^1$Steward Observatory,
  University of Arizona, 933 North Cherry Avenue, Tucson, AZ 85721, USA \\
  $^2$Observatories of the Carnegie Institution for Science, 813 Santa
  Barbara St, Pasadena, CA 91101, USA \\ $^3$Center for Astrophysics
  and Space Science, University of California San Diego, La Jolla, CA
  92093, USA}

\begin{document}
\date{Accepted 0000, Received 0000, in original form 0000}
\pagerange{\pageref{firstpage}--\pageref{lastpage}} \pubyear{2002}
\def\arcdeg{\degr}
\maketitle
\label{firstpage}

\begin{abstract}

  We present low-resolution near-infrared (IR) 0.8-2.5 $\mu$m spectra
  of Supernova (SN) 2009ip, taken immediately before, during, and just
  after its rapid brightening in late September/October 2012.  The
  first epoch shows the same general spectral characteristics as the
  later epochs (smooth continuum, narrow H and He~{\sc I} emission
  lines), but the IR continuum shape is substantially redder than the
  later epochs.  The epoch 1 continuum can be approximated by
  reddening the peak-luminosity (epoch 3) spectrum by $E(B-V)$=1.0
  mag, but the blue color seen in visual-wavelength spectra at the
  same time indicates that strong wavelength-dependent extinction by
  circumstellar dust is not the correct explanation.  Instead, we
  favor the hypothesis that the redder color before the brightening
  arises from excess emission from hot $\sim$2000~K circumstellar
  dust.  The minimum radius ($\ga$120 AU) deduced from the dust
  temperature and observed luminosity of the transient, combined with
  the observed expansion speed in the precursor outbursts of
  SN~2009ip, is consistent with an ejection at least 1.1 yr earlier.
  The mass of hot dust indicated by the IR excess is
  $\sim$4$\times$10$^{-7}$ $M_{\odot}$, although this is only a lower
  limit since the near-IR data do not constrain the mass of cooler
  dust.  Thus, the observed pre-SN outbursts of this object were able
  to efficiently form dust into which the SN ejecta and radiation now
  propagate.  This is consistent with the notion that the same pre-SN
  eruptions that generally give rise to SNe~IIn also give rise to the
  dust needed for their commonly observed IR echoes.  We also discuss
  some aspects of the IR line profiles, including He~{\sc i}
  $\lambda$10830.

\end{abstract}

\begin{keywords}
  circumstellar matter --- stars: variables: other --- stars: winds,
  outflows --- supernovae: general --- supernovae: individual (SN
  2009ip)
\end{keywords}

\section{INTRODUCTION}

Supernova (SN) 2009ip provided a truly remarkable sequence of events,
first noted as a SN ``impostor'' or eruption of a luminous blue
variable (LBV) in 2009 (Smith et al.\ 2010), and then subsequently
observed as a core-collapse SN in mid/late 2012 (Mauerhan et al.\
2013).  It joins SN~1987A (Walborn et al.\ 1989; Rousseau et al.\
1978) as the only SN for which we have a direct detection {\it and a
  spectrum} of the progenitor star.  In the case of SN~2009ip,
however, we have a much more detailed record and higher quality data
pertaining to its unstable pre-SN state.  It is reminiscent of the
pre-SN outburst seen 2 yr before SN~2006jc (Pastorello et al.\ 2007),
but in that case only one outburst was detected and no spectrum was
obtained.  A brief recounting of SN~2009ip's pre-SN activity is as
follows:

After the discovery of SN~2009ip as a new transient source in August
2009 (Maza et al.\ 2009), it was recognized that it was not a true SN,
but was instead a non-terminal eruptive transient similar to LBVs.
Smith et al.\ (2010) presented archival {\it Hubble Space Telescope}
({\it HST}) images, archival ground-based photometry, and new
photometry and spectra of the outburst. {\it HST} data revealed that
10 yr prior to discovery, the quiescent progenitor was a luminous
supergiant with $L \simeq 10^6 L_{\odot}$ and a likely initial mass of
around 50-80 $M_{\odot}$, while the archival ground-based photometry
documented a $\sim$5 yr long S Dor-like LBV phase that preceded the
2009 discovery.  The 2009 outburst itself had a peak absolute
magnitude of around $-$14.5 (Smith et al.\ 2010), similar to giant LBV
eruptions (Smith et al.\ 2011).  Spectra of the 2009 outburst showed
strong Balmer emission lines similar to LBV eruptions (e.g., Smith et
al.\ 2011), indicating an expansion speed of order 550 km s$^{-1}$.
The 2009 outburst of SN~2009ip was unusual compared to most LBV-like
eruptions, however, due to the rapid brightening and fading over a few
weeks instead of a few months to years (Smith et al.\ 2011).  Foley et
al.\ (2011) presented additional spectra and a reanalysis of the
archival {\it HST} data that reinforced these conclusions. After these
first papers on the 2009 outburst, SN~2009ip then had another outburst
of similar magnitude in 2010 (Drake et al.\ 2010), and it has recently
been reported (Pastorello et al.\ 2013) that it had additional
eruptions of similar brightness since then as well.

On 2012 July 24 (UT dates are used throughout this paper), SN~2009ip
was discovered to have yet another outburst (Drake et al.\ 2012).
This new outburst was of similar magnitude to the previous ones, and
at first it was assumed that this was yet another LBV-like eruption.
Foley et al.\ (2012) reported that an early spectrum obtained on 2012
Aug.\ 24 was consistent with spectra of the previous LBV-like
outbursts, with line widths around 600 km s$^{-1}$.  However, spectra
obtained later on 2012 Sep.\ 7-16 (Smith \& Mauerhan 2012) revealed a
transformation in the spectrum of SN~2009ip, which now showed very
broad Balmer emission and absorption lines out to speeds of 13,000 km
s$^{-1}$.  These broad features indicated that the impostor SN~2009ip
was becoming a true core-collapse SN (Mauerhan et al.\ 2012), since
these spectral features had never been seen previously in this object
or any LBV eruption.  Pastorello et al.\ (2013) have pointed out that
broad features were seen previously, but these were only seen in
absorption (as noted earlier by Smith et al.\ 2010 and Foley et al.\
2011), while emission-line components of SN~2009ip had always shown
narrow or intermediate-width emission lines prior to Sep.\ 2012.
Small amounts of very fast material have been seen around LBVs, as in
the case of $\eta$~Carinae (Smith 2008), but the bulk velocity of most
of the ejected mass in LBV eruptions is much slower.

\begin{figure}\begin{center}
\includegraphics[width=3.1in]{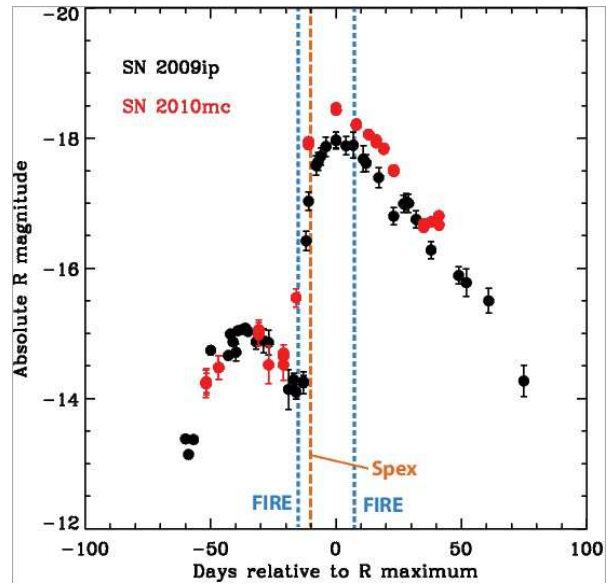}
\end{center}
\caption{Absolute $R$-band magnitude light curve of SN~2009ip (black
  points) from Mauerhan et al.\ (2012; but with the three earliest
  points taken from Pastorello et al.\ 2013), showing the dates of our
  FIRE and Spex spectra relative to the time of maximum luminosity
  (see Table 1).  The red points show the recently published light
  curve of SN~2010mc from Ofek et al.\ (2013a), which appears to have
  an almost identical luminosity evolution (i.e., the absolute
  $R$-band magnitude of SN~2010mc was not scaled).}
\label{fig:lc}
\end{figure}

The broad emission lines in SN~2009ip were first noted in Sep.\ 2012,
around the time of peak luminosity for the precursor outburst (see
Figure~\ref{fig:lc}).  The object faded over the next week or so, but
then brightened very suddenly, rising $\sim$3 mag in a few days.  This
very rapid brightening was probably caused by the fast SN ejecta
crashing into slower circumstellar material (CSM) ejected by the
precursor eruptions (Mauerhan et al.\ 2013).  Prieto et al.\ (2013)
presented exceptionally high-cadence photometry of this very rapid
brightening phase, also concluding that this was most likely due to
the onset of SN-CSM interaction similar to other SNe IIn.  Indeed, as
it reached peak luminosity, the spectrum took on the character of a
normal SN~IIn with a smooth blue continuum and Balmer lines exhibiting
narrow line cores and Lorentzian scattering wings (Mauerhan et al.\
2012).  Levesque et al.\ (2013) have recently presented additional
spectra of the SN, highlighting aspects of its Balmer
decrement. SN~2009ip gives vivid confirmation of the idea that very
massive and eruptive LBV-like stars can explode to produce SNe~IIn
(e.g., Smith et al.\ 2007, 2008a, 2011; Gal-Yam et al.\ 2007; Gal-Yam
\& Leonard 2009).

In Figure~\ref{fig:lc} we also plot the light curve of SN~2010mc for
comparison, recently reported by Ofek et al.\ (2013).  Ofek et al.\
discussed the remarkable detection of a precursor outburst in this
Type~IIn event as well, but they did not note the uncanny similarity
to the light curve of SN~2009ip (which was still underway as their
paper was in preparation).  Figure~\ref{fig:lc} emphasizes that {\it
  these two objects are nearly identical} in terms of peak
luminosities and timescales for both the precursor and SN outbursts
(the absolute magnitudes have not been adjusted).  Since most SNe~IIn
observed thus far have been discovered near peak and lack high-quality
pre-SN observations, it may be the case that the erratic pre-SN
variability of SN~2009ip is not so unusual.

Here we present near-IR spectra of the remarkable 2012 explosion of
SN~2009ip.  We discuss implications for the dusty environment produced
by the precursor LBV eruptions, as well as aspects of the line
profiles.

\begin{table}\begin{center}\begin{minipage}{3.3in}
      \caption{Infrared Spectroscopy of SN~2009ip}\scriptsize
\begin{tabular}{@{}lcccc}\hline\hline
  Date        &Day   &Tel./Instr.  &R  &$\Delta\lambda$   \\ 
  (y-m-d)     &(max) &             &       &($\mu$m)    \\   \hline
2012-09-22  &-15     &Magellan/FIRE     &300-500  &0.8-2.5     \\
2012-09-27  &-10     &IRTF/Spex         &120      &0.7-2.5     \\
2012-10-14  &+7      &Magellan/FIRE     &300-500  &0.8-2.5     \\
\hline
\end{tabular}\label{tab:pcyg}
\end{minipage}\end{center}
\end{table}

\section{OBSERVATIONS}

\begin{figure*}\begin{center}
\includegraphics[width=5.9in]{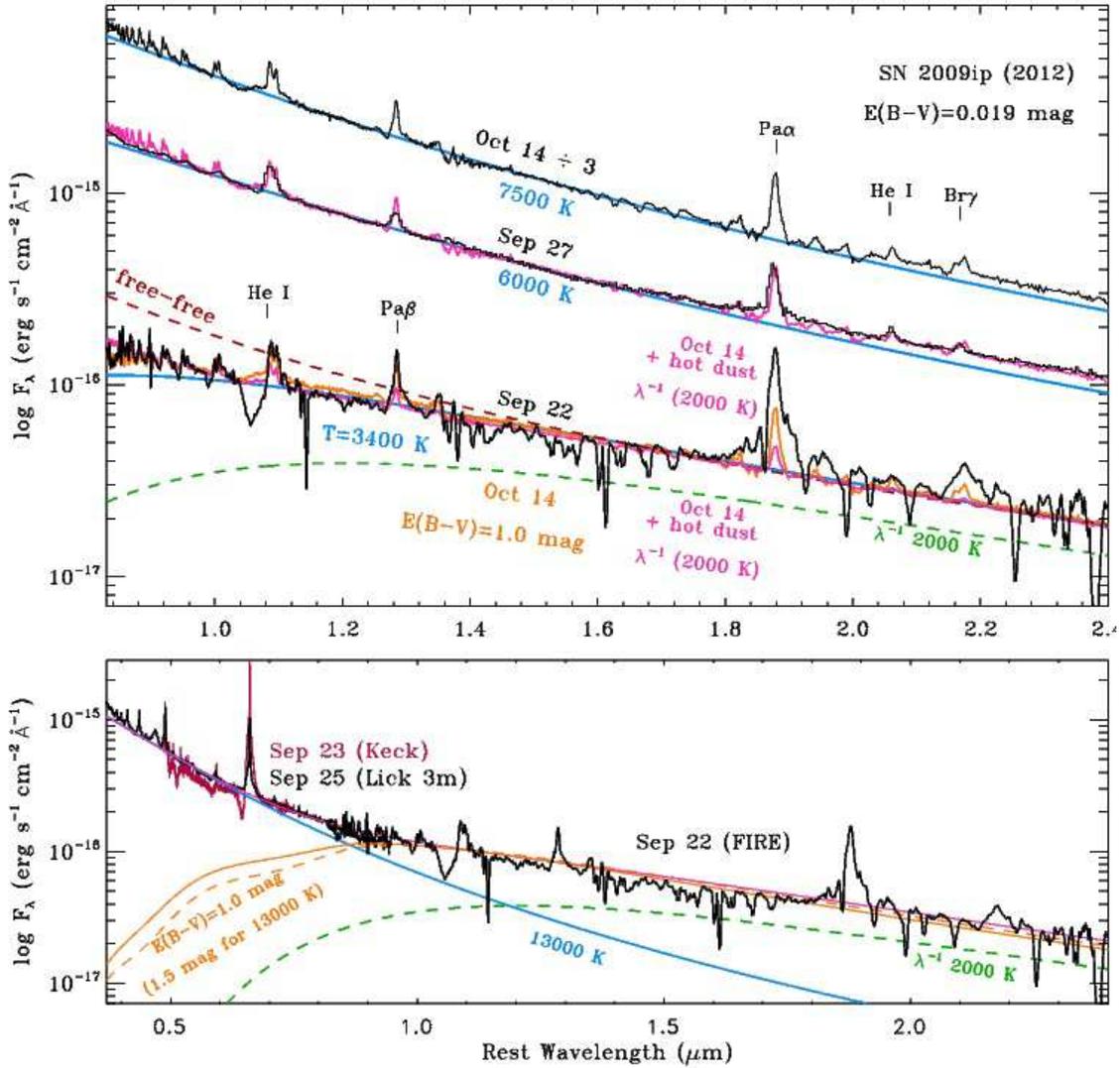}
\end{center}
\caption{(Top) Near-IR spectra of SN~2009ip during the 2012
  brightening.  The black spectra are taken with Spex and FIRE on the
  dates indicated (see Table 1), corrected for Galactic reddening of
  $E(B-V)$=0.019 mag (Smith et al.\ 2010).  The colored curves show
  various possible interpretations to match the continuum shape: (1)
  the blue curves show representative blackbodies; (2) the orange
  spectrum shows the Oct.\ 14 spectrum (near peak luminosity) reddened
  by $E(B-V)$=1.0 mag to match the continuum shape of the pre-SN
  spectrum on Sep.\ 22; (3) the two magenta spectra once again use the
  Oct.\ 14 spectrum, but this time the continuum shape is augmented by
  a 2000~K greybody with $\lambda^{-1}$ emissivity, giving a rough
  example of how hot dust might add to the spectrum (the dust
  contribution is shown alone by the green dashed curve).  The magenta
  spectrum matched to the Sep.\ 27 spectrum has less relative
  contribution from hot dust, and relatively more from the scaled
  Oct.\ 14 spectrum.  (Bottom) The same Sep.\ 22 IR spectrum as in the
  top panel, but plotted alongside nearly simultaneous
  visual-wavelength spectra taken 1-3 days later, from Mauerhan et
  al.\ (2013).  The shape of the combined optical/IR continuum is best
  explained by a 13,000 K blackbody (blue) plus the same 2000 K dust
  component as in the top panel (green dashed; the magenta curve is
  the addition of the two components). Also shown for comparison are a
  7500 K blackbody reddened by E(B-V)=1.0 mag (orange solid curve, as
  in the top panel) and a hotter 13,000 K blackbody reddened by 1.5
  mag (orange dashed).}
  \label{fig:irspec}
\end{figure*}

We obtained three IR spectra of SN~2009ip, which were taken immediately
before its fast rise in brightness, during the rise, and around the
time of peak luminosity.  Figure~\ref{fig:lc} shows the three epochs
of our spectroscopy relative to the light curve of SN2009ip's 2012
outburst from Mauerhan et al.\ (2012).  Observational information is
listed in Table 1, and the resulting spectra are plotted in
Figure~\ref{fig:irspec}.  All spectra in Figure~\ref{fig:irspec}
(plotted in black) have been corrected for an assumed Galactic
reddening value of $E(B-V)$=0.019 mag, following Smith et al.\ (2010).

Our first near-IR spectrum was obtained on 2012 Sep.\ 22 (UT) using
the Folded-port InfraRed Echellette spectrograph (FIRE; Simcoe et al.\
2008, 2010) on the 6.5-m Magellan Baade Telescope. This was only 2
days prior to the dramatic rapid rise in brightness of SN~2009ip.  The
low-dispersion, high-throughput prism mode provides a spectrum that
spans a wavelength range of 0.8$-$2.5 $\mu$m at a resolving power
$\lambda/\Delta\lambda$ = 300$-$500.  We completed an ABBA sequence on
SN~2009ip, with an average airmass of 1.16, and immediately afterward
obtained a spectrum of an A0~V standard star (HIP~113376) for flux
calibration and telluric correction, as described by Vacca, Cushing,
\& Rayner (2003). Data were reduced using the FIREHOSE pipeline.
Portions of this spectrum were presented and discussed by Ofek et al.\
(2013b) regarding relative line velocities.  However, we noted that
there may have been thin cirrus clouds during the observation of the
standard star, which might affect the absolute flux calibration;
because of this and possible slit losses that are difficult to
quantify, the resulting flux-calibrated spectrum appeared brighter
than indicated by simultaneous photometry, so the spectrum is scaled
down in Fig.~\ref{fig:irspec}.

A second epoch of low-resolution near-IR spectra for SN~2009ip were
obtained with the SpeX spectrograph (Cushing et al.\ 2004) on the NASA
Infrared Telescope Facility (IRTF) on 2012 Sep.\ 27 (UT) in clear
conditions with 0$\farcs$5 seeing.  This spectrum was taken during the
rapid rise, well before SN~2009ip had reached its peak luminosity, and
an initial report was given by Burgasser et al.\ (2012).  The
prism-dispersed mode and 0$\farcs$5 slit (aligned with parallactic)
were used, providing 0.7--2.5~$\micron$ coverage with resolution
$\lambda$/$\Delta\lambda$ $\approx$ 120, and dispersion of
20--30~{\AA}~pixel$^{-1}$.  Eight exposures were obtained in an ABBA
pattern at an average airmass 1.52, for a total exposure of 840~s.
The nearby A0~V star HD~202941 ($V$ = 7.05 mag) was observed
immediately after for flux calibration and telluric absorption
correction, along with internal flat field and arc lamp exposures.
Data were reduced with the IDL SpeXtool package, version 3.4 (Vacca et
al.\ 2003; Cushing et al.\ 2004), using standard settings.

Our third epoch was obtained on 2012 Oct.\ 14 (UT), once again using
FIRE on Magellan.  The low-resolution instrumental setup, observations
of an A0 V standard, and data reduction procedures were nearly the
same as described above for the previous FIRE epoch, except that we
used a different standard star (HD~219341).

\section{DUST AND THE IR CONTINUUM SHAPE}

A striking aspect of the spectral sequence in Figure~\ref{fig:irspec}
(top) is that the first-epoch spectrum on Sep.\ 22 is substantially
different from those on Sep.\ 27 and Oct.\ 14, which appear
qualitatively very similar to each other.  One particular difference
is in the continuum shape.  For comparison, representative blackbody
curves (blue) are plotted along with each epoch of the observed
spectra, suggesting that the apparent color temperature increases from
$\sim$3400~K on Sep.\ 22, to 6000~K and 7500~K on the two later
epochs.  Continuum slopes matching temperatures of 6000-7500~K are
typical for SNe~IIn, and the spectral characteristics (strong H lines)
are consistent with that range of temperatures.  However, the much
lower temperature of 3400~K implicated by the IR continuum slope on
Sep.\ 22 is not commensurate with the presence of He I and H emission
lines in the spectrum.  Therefore, a cooler photospheric temperature
is not the likely explanation for the continuum shape (the 3400~K
blackbody underpredicts the flux near 1 $\mu$m anyway).  Two possible
alternatives involving dust are discussed below.

First, a redder color on Sep.\ 22 than at later times could
potentially arise from significant reddening by CSM dust, which is
destroyed at later epochs as the SN brightens and the shock expands.
The orange curve in Figure~\ref{fig:irspec} (top panel) shows the
Oct.\ 14 spectrum reddened by $E(B-V)$=1.0 mag, providing a
satisfactory match to the shape of the observed spectrum on Sep.\ 22.
This is unlikely to be the real explanation for the observed color,
however, because the amount of reddening required to match the IR
continuum shape would cause even more severe reddening at visual
wavelengths.  This is demonstrated in the bottom panel of
Figure~\ref{fig:irspec}, where we plot visual-wavelength spectra taken
1-3 days later (already published by Mauerhan et al.\ 2013) alongside
the Sep.\ 22 FIRE spectrum.  The two orange curves (solid and dashed)
show hot blackbodies (7500 and 13,000 K, respectively) that are
reddened to approximate the shape of the near IR continuum, using
$E(B-V)$=1.0 and 1.5 mag, respectively.  While these reddened
blackbodies seem to match the IR continuum shape, the flux at visual
wavelengths fall far below the observed visual wavelength spectra
taken only 1 and 3 days later (note that these visual spectra have
been adjusted in flux by a few per cent so that the $\sim$9000 \AA\
fluxes match the IR spectra, correcting for the fact that SN~2009ip
was variable).

Instead of reddening, a more likely explanation for the IR color is an
extra contribution from {\it emission} of dust instead of absorption.
The magenta spectrum plotted against the observed Sep.\ 22 spectrum
is, once again, the spectrum near peak on Oct.\ 14, but with an added
contribution from hot dust emission (Fig.~\ref{fig:irspec}, top
panel).  For simplicity, we used a 2000~K blackbody modified by
$\lambda^{-1}$ emissivity.  When added to the Oct.\ 14 spectrum, it
matches the overall shape of the Sep.\ 22 near-IR continuum very well.
The bottom panel of Figure~\ref{figLirspec} shows that the combination
(magenta curve) of a hot blackbody (13000 K, blue) and the same 2000 K
dust component (green dashed) can match both the visual wavelength and
IR spectra simultaneously.  The fact that the visual-wavelength
continuum requires a hotter blackbody (13,000) than the temperature
implied by fitting the IR continuum alone (7500K) merely reflects that
fact that IR wavelengths on the Rayleigh-Jeans tail of a Planck
function are a less sensitive probe of the underlying hot component.

This IR excess presumably arises from CSM dust heated by the
luminosity of the precursor transient source, as in an ``IR
echo''. Dust at around this temperature is as hot as dust can be,
residing at the innermost radius of the dust shell, inside of which
the dust is destroyed.\footnote{Note that a high dust temperature near
  2000~K does not necessarily indicate C-rich dust, as is often
  assumed.  Dust formed in the colliding-wind shock of $\eta$ Carinae
  has a similar high temperature (Smith 2010) even though $\eta$ Car
  is known to be severely carbon deficient.  For $\eta$ Car, the hot
  dust is thought to be corundum (Al$_2$O$_3$), which condenses at a
  similar high temperature.}  This general picture is similar to the
IR echoes observed in many SNe, particularly common among SNe~IIn (see
Fox et al.\ 2011, and references therein) due to their dense and dusty
pre-existing CSM.  On Sep.\ 22 before the rapid brightening, SN~2009ip
had a luminosity of roughly $L\,\simeq\,$4$\times$10$^7$ $L_{\odot}$.
CSM dust in equilibrium at 2000~K would then reside at a radius of

\begin{equation}
  R \, \simeq \, 120 \, AU \, \Big{(}\frac{Q \, L}{10^{7.6} 
    L_{\odot}}\Big{)}^{\frac{1}{2}} \Big{(}\frac{T_d}{2000 K}\Big{)}^{-2}
\end{equation}

\noindent (where $Q$ is the ratio of the grain absorption to emission
efficiency), which is around 120 AU for large grains that behave like
blackbodies ($Q \simeq 1$), or somewhat farther from the star for
smaller grain sizes that have larger values of $Q$.  Equation (1)
really provides a lower limit to the radius of the dust shell, since
it could be farther away if dust is condensing as an ejected shell
expands and cools (e.g., Kochanek et al.\ 2011).  This is, therefore,
probably not dust that formed from the beginning of the 2012 eruption
$\sim$40 days earlier, because that material would be too close to the
star.  Expanding at around 550 km s$^{-1}$ (Smith et al.\ 2010), the
implied age is at least 1.1 yr, making it seem quite plausible that
this dust formed from ejecta in one of its previous documented
outbursts (see Mauerhan et al.\ 2012, Smith et al.\ 2010; Drake et
al.\ 2010; Pastorello et al.\ 2013).\footnote{In this context, it is
  interesting to note that Foley et al.\ (2011) reported a similar
  $\sim$2100 K dust excess around the LBV-like transient UGC2773-OT.}
In any case, it should be noted that this hot CSM dust that
contributes to the emission in the IR spectrum on Sep.\ 22 will not
necessarily cause any corresponding absorption along our line-of-sight
to the SN photosphere, since the geometry may be globally
non-spherical or clumpy.

The contribution of this dust emission could weaken as time proceeds,
either because some of the nearby CSM dust gets destroyed, or because
the dust emission simply makes a smaller contribution to the total
flux as the brightness of the SN photosphere increases.\footnote{Note
  that our first spectrum was taken just 2 days before the rapid
  brightening, when the optical flux had faded by about 1 mag from the
  peak of the precursor event (see Fig.~\ref{fig:lc}).  This drop in
  flux from the underlying photosphere would serve to enhance the
  contrast of the IR excess from dust at this epoch.}  Both appear to
be happening.  By the time SN~2009ip reached its peak, it had a
luminosity of roughly 1.3$\times$10$^9$ $L_{\odot}$.  The same dust
that was at $R$=120 AU and 2000~K would then be heated to an
equilibrium temperature of 4800~K (equation 1), vaporizing dust grains
at that radius.  (Moreover, the blast wave of the SN itself may have
been approaching roughly the same radius by this point.)  This
destruction should happen as the SN brightens, but the IR excess might
not disappear if there is additional dust at larger radii from
previous outbursts that could be heated to the same temperature by the
increasing luminosity.  In fact, some IR excess emission does seem to
persist.  The magenta spectrum plotted against the Sep.\ 27 spectrum
in Fig.~\ref{fig:irspec} (top) shows the same contribution of hot dust
at the same 2000~K temperature, except that the relative contribution
of the dust is less.  This smaller relative contribution from hot dust
matches the spectral shape very well, especially the excess flux
longward of 1.8 $\mu$m, and mimics the appearance of a cooler
photospheric temperature (6000~K vs. 7500~K at peak).  A smaller
contribution of hot dust may also be present even in the Oct.\ 14
spectrum around the time of peak luminosity, since it appears to have
a small amount of excess flux beyond 2 $\mu$m.

The mass of hot dust can be inferred from the IR excess if the grains
are small ($a\la0.2 \ \mu$m; following Smith \& Gehrz 2005 and Smith
et al.\ 2008b), which for grains with a density of $\rho$=2.25 g
cm$^{-3}$ can be expressed as

\begin{equation}
  M_d \approx 5 \times 10^{-7} \, M_{\odot} \, \Big{(} \frac{L_d}{10^7 \, L_{\odot}} \Big{)}  \Big{(} \frac{T_d}{2000 \, K} \Big{)}^{-6} 
\end{equation}

\noindent where $L_d$ is the luminosity of the hot dust component at a
temperature $T_d$.  The 2000 K dust that contributes the IR excess in
our pre-SN spectrum on Sep 23 has a luminosity $L_d \approx
9\times$10$^6$ $L_{\odot}$, although this comes with a large
uncertainty of perhaps $\pm$50\% due to possible slit losses in the
exposures of SN~2009ip or the standard star.  Nevertheless, this
provides a very rough estimate of the mass of hot dust of roughly
(3-6)$\times$10$^{-7}$ $M_{\odot}$.  This is, of course, only a lower
limit to the total circumstellar dust mass because our near-IR data
only trace the hottest dust, while a large additional mass of cooler
dust may emit at longer IR wavelengths.

Lastly, we note that additional IR flux from free-free emision due to
an ionized stellar wind, which often causes IR excess in hot stars
(Wright \& Barlow 1975), cannot explain the red color because the
slope from free-free emission is too blue at the shortest IR
wavelengths.  The dashed brown curve in Figure~\ref{fig:irspec} (top)
shows the expected free-free IR emission from a constant-speed stellar
wind.

\begin{figure}\begin{center}
\includegraphics[width=2.6in]{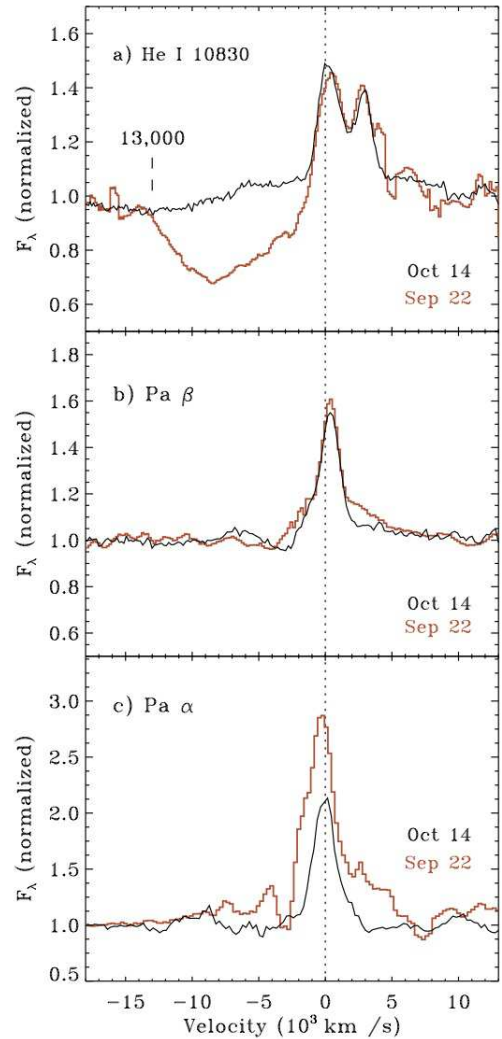}
\end{center}
\caption{Individual line profiles before the sharp rise and soon after
  peak luminosity.  (a) He~{\sc i} $\lambda$10830, (b) Pa $\beta$, and
  (c) Pa $\alpha$.}
  \label{fig:vel}
\end{figure}

\section{LINE PROFILES}

Line profiles for the three brightest lines in the IR spectrum are
plotted in Figure~\ref{fig:vel}, showing ``before'' and ``after''
spectra of He~{\sc i} $\lambda$10830, Pa~$\beta$, and Pa~$\alpha$
taken at the same resolution with the FIRE spectrograph.

The most notable change in the line profiles is seen in He~{\sc i}
$\lambda$10830.  In the first epoch before the brightening, this line
exhibits a very strong and broad P Cygni absorption trough extending
to a blue edge of $-$13,000 km s$^{-1}$.  This is the same speed as
seen in the blue edge of the P Cygni absorption in Balmer lines over
the preceding week or so (Mauerhan et al.\ 2012).  This broad He~{\sc
  i} absorption then disappears when the SN brightens to
peak.\footnote{Note that some significant portion (20$-$50\%,
  depending on wavelength) of the near-IR continuum comes from hot CSM
  dust in an echo, rather than from the SN itself.  This has the
  effect of diluting the apparent relative strength of
  emission/absorption lines.  In other words, the IR
  emission/absorption features are stronger than they appear in this
  spectrum (i.e. the true intrinsic equivalent width is larger).}
Balmer lines show the same disappearance of the broad profiles at peak
(Mauerhan et al.\ 2012), presumably for the same reason -- i.e., the
spectrum becomes dominated by opaque CSM interaction.  The narrow
emission components of He~{\sc i} $\lambda$10830 show little or no
change before and after the rapid brightening, relative to the
continuum.  The metastable nature of He~{\sc i} $\lambda$10830
absorption makes it difficult to use it to draw conclusions about the
relative He abundance in the fast ejecta, and there are few SNe~IIn
with comparable near-IR spectra for comparison.

Pa~$\beta$ and Pa~$\alpha$ do not show any strong broad P Cygni
absorption, as is seen in Balmer lines and He~{\sc i}
$\lambda$10830. There is a relatively faint underlying broad emission
component to both Paschen lines, appearing as broad wings that weaken
or disappear at the last epoch.  The broad component makes a much
stronger relative contribution to Pa~$\alpha$ than to Pa~$\beta$,
which is the opposite of what is expected for electron scattering,
where higher-order Paschen lines formed at higher optical depth have
stronger electron scattering wings (Dessart et al.\ 2009).  There is
an apparent P Cygni absorption trough at roughly $-$3000 km s$^{-1}$
in Pa~$\alpha$ (with a width that is similar to the spectral
resolution of our instrument), but it is difficult to judge the
reality of this feature because of the strong telluric absorption in
the vicinity of Pa~$\alpha$.  As with He~{\sc i} $\lambda$10830, the
narrow components of the Paschen lines show little or no change with
time, compared to the continuum, since they arise primarily in the
CSM.

\smallskip\smallskip\smallskip\smallskip
\noindent {\bf ACKNOWLEDGMENTS}
\smallskip
\footnotesize

We thank an anonymous referee for helpful comments. MMK acknowledges
generous support from the Hubble Fellowship and Carnegie-Princeton
Fellowship.

{\it Facilities:} Magellan:Baade (FIRE), IRTF (Spex).

\end{document}